\documentclass[twocolumn,showpacs,preprintnumbers,amsmath,amssymb]{revtex4-1}

\usepackage{graphicx}
\usepackage{amssymb}
\usepackage{hyperref}

\begin{document}

\title{Physical properties of CeIrSi with trillium lattice frustrated magnetism}

\author{F. Kneidinger$^1$, I. Zeiringer$^2$, A. Siderenko$^1$
E. Bauer$^1$, H. Michor$^1$, P. Rogl$^2$,  J.G. Sereni$^3$}

\address{$^1$ Institute of Solid State Physics, TU Wien, A-1040 Wien, Austria\\
$^2$ Institut fuer Physikalische Chemie, Universitaet Wien, A-1020 Wien, Austria\\
$^3$ Low Temperature Division CAB-CNEA, CONICET, 8400 S.C. de
Bariloche, Argentina}

\begin{abstract}

{Magnetic ($\chi$), transport ($\rho$) and heat capacity ($C_m$)
properties of CeIrSi are investigated to elucidate the effect of
geometric frustration in this compound with trillium type
structure because, notwithstanding its robust effective moment, $\mu_{\rm
eff}\approx 2.46\mu_B$, this Ce-lattice compound does not undergo
a magnetic transition. In spite of that  it shows broad
$C_m(T)/T$ and $\chi(T)$ maxima centered at $T_{max}\approx
1.5$\,K, while a $\rho \propto T^2$ thermal dependence, 
characteristic of electronic spin coherent fluctuations, is observed below
$T_{coh} \approx 2.5$\,K. Magnetic field does not affect
significantly the position of the mentioned maxima up to $\approx
1$\,T, though $\chi(T)$ shows an incipient structure that
completely vanishes at $\mu_0 H \approx 1$\,T. Concerning the
$\rho \propto T^2$ dependence, it is practically not affected by magnetic
field up to $\mu_0 H = 9$\,T, with the residual resistivity $\rho_0(H)$ slightly decreasing and $T_{coh}(H)$ increasing. These results are compared with the physical properties observed in other frustrated 
intermetallic compounds.}

\end{abstract}
\date{\today}

\maketitle

\section{Introduction}

The lack of magnetic order in lattice arrangements of robust
magnetic moments allows to access to exotic ground states with
high density of low energy excitations. Two typical scenarios allow
to prevent the development of magnetic order: i) the weakness of the
magnetic interactions or ii) the frustration of antiferromagnetic
interactions. In the former, cerium magnesium nitrate hydrate
(CMN) is the exemplary case because it remains 
paramagnetic down to $\approx$ 2\,mK due to the large Ce-Ce
spacing $d_{Ce-Ce}\approx 11$\,\AA \cite{CMN} 
and the absence of conduction electrons. In the latter
context, two circumstances may produce frustration; one due to
geometrical constraints like triangular (2D) or tetrahedral (3D)
spin lattices, or because of the competition between nearest (nn)
and next nearest neighbors (nnn) interactions \cite{geomFrustr}.
The pyrochlore structure of the Dy$_2$Ti$_2$O$_7$ spin-ice
\cite{Dy2Ti2O7} is an exemplary system for the
3D-tetrahedral coordination case, whereas some 2-2-1 \cite{CePdAgIn} compounds 
showing a network of triangles and squares exhibit magnetic frustration in their
basal (2D) planes. Finally, the competition between 'nn' and 'nnn' interactions
can be exemplified by Yb$_4$LiGe$_4$ \cite{Yb4LiGe4}.

Among the crystalline structures favoring 3D geometric frustration, the cubic
trillium (LaIrSi-type) structure \cite{Klepp,Chevalier,Monatshefte} should provide
an ideal  playground for a study of the competition between RKKY interactions and 
frustration effects. Several light rare earth - iridium - silicides (REIrSi, RE = rare earth) 
are members of this structure type. 

Ternary intermetallics REIrSi  have been intensely studied 
in the past few decades. The respective  
crystal structure depends on the distinct rare earth element. 
Compounds with RE = La, Ce, Pr and Nd exhibit the cubic LaIrSi structure type 
(space group $P2_13$) \cite{Chevalier,Klepp}, where inversion symmetry is missing. 
This structure type is a ternary ordered version of the binary $\rm SrSi_2$ type
(space group $P4_132$). Due to ordering of the Ir and Si atoms in LaIrSi, symmetry
is lowered, as evidenced from the respective space groups. 
The iridium and silicon atoms build up a three-dimensional [IrSi] network with rather 
short Ir-Si distances, inferring strong covalent bonding \cite{Monatshefte}. On the
other hand,  distances of between rare earth ions and Si or Ge are much larger,
evidencing a much weaker bonding \cite{Monatshefte}.

Silicides with heavy rare earth elements (RE from Gd to Lu), 
however, crystallizing in the orthorhombic TiNiSi 
structure (space group $Pnma$) \cite{Shoemaker}. 
The same is true for ScIrSi and YIrSi.
In this structure, the  iridium  and  silicon
atoms form a three-dimensional [IrSi] network in which the
heavy rare earth  atoms are located in distorted hexagonal channels. 
Short Ir-Si distances
are indicative for strong Ir-Si bonding \cite{Mishra}. 
The Sm based compound with an empirical formula $\rm SmIr_{0.266}Si_{1.734}$ 
is found in the tetragonal $\alpha$-ThSi$_2$ structure type, space group $I4_1/amd$. 

For LaIrSi, a superconducting phase transition at $T_c = 2.3$~K 
has been obtained in Ref. \cite{Chevalier}.
A subsequent study by Evers et al. \cite{Evers} revealed 
superconductivity below  $T_c = 1.5$~K. 
Upon annealing, this temperature shifted up to 2.3~K. The authors, 
however, concluded from just a small anomaly
at $T = T_c$ that superconductivity in this compound 
is not a bulk property.  
The authors of the present study (compare Ref. \cite{Kneidinger}) also 
have not observed bulk superconductivity,
as evidenced from from missing an appropriate jump in the heat capacity data and
from the non-zero resistance at 350~mK.  

While CeIrSi was characterised from temperature dependent susceptibility data 
as a paramagnet without magnetic ordering down to 1.5~K \cite{Monatshefte}, 
for NdIrSi, Chevalier et al. \cite{Chevalier} reported 
a ferromagnetic ground state below $T_C = 10$~K from a spontaneous magnetisation.
This, in addition, was supported from a positive value of the 
paramagnetic Curie temperature ($\theta _p = 12$~K). 
The study by Heying et al. \cite{Monatshefte} confirmed the LaIrSi crystal structure 
for PrIrSi, too, but no physical properties were reported. 

Magnetic properties and magnetic structures of REIrSi 
(RE from Tb to Er) were revealed by Szytula et al. 
\cite{Szytula}. Sine modulated and collinear antiferromagnetic 
orders at lower temperatures have been 
derived in this study from elastic neutron scattering experiments. 

Within  the  series REIrSi  (RE = Gd,  Ho,  Er,  Yb, Lu),
the unit cell volume decreases monotonically from  GdIrSi  to  YbIrSi \cite{Monatshefte}.
This would infer the magnetic $4f^{13}$ electronic configuration of the Yb ion in 
this ternary compound. 
Thus, paramagnetic behaviour is  expected \cite{Mishra}. A temperature independent 
susceptibility evidences Pauli paramagnetism for LuIrSi \cite{Monatshefte}.

In the present study we aim a thorough characterisation of CeIrSi in the context of
the LaIrSi crystal structure. Because of distinct features
of the trillium structure, a weaker Ce-Ce connectivity
with the next nearest neighbours is expected, compared to cases with 
the pyrochlore structure, as it forms a
three-dimensional network of corner-sharing triangles (resembling
a trillium flower, see Fig.~\ref{F1}) instead of corner-sharing
tetrahedra. Within this peculiar structure the 6
Ce nearest neighbours  are at $d_{Ce-Ce}= 3.855(1)$\,\AA \cite{Monatshefte}, which
is close to a direct Ce-Ce contact. In this work structural,
thermodynamic and transport properties of CeIrSi are investigated
to elucidate the effect of geometric frustration 
with respect to ground state properties of this compound.

\section{Experimental details}

\subsection{Sample preparation and characterization}

Polycrystalline samples of LaIrSi and CeIrSi were 
obtained by melting pure ingots of respective components  
weighted in proper stoichimetric composition 
in an arc furnace with argon atmosphere, using Ti as getter 
material. Several remelting processes were
carried out to assure sample homogeneity. Subsequently, the samples were
sealed in quartz tubes and annealed for one weak. X-ray powder
diffraction (XRD), scanning electron microscopy (SEM) and electron
probe micro analysis (EPMA) were used for the characterization of the samples. 

As a derivative of the non-centrosymmetric SrSi2-type 
structure \cite{Klepp}, LaIrSi and CeIrSi compounds were found to 
have respective lattice parameters: $a =$ 6,3766(3)\,\AA  and 6,2951(1)\,\AA. 
The actual relative concentration was determined by EPMA 
as: 33.4; 34.7; 31.9\% for LaIrSi and 34.3; 32.7; 33.0\% 
for CeIrSi. The LaIrSi sample contains small amounts ($<2$\%)
of LaIr$_2$Si$_2$ as an impurity phase, and CeIrSi also 
contains about 2\% of CeIr$_2$Si$_2$, CeSi$_{1.7}$ and 
small amounts of cerium oxide as impurities.

\begin{figure}[tb]
\begin{center}
\includegraphics[width=18pc]{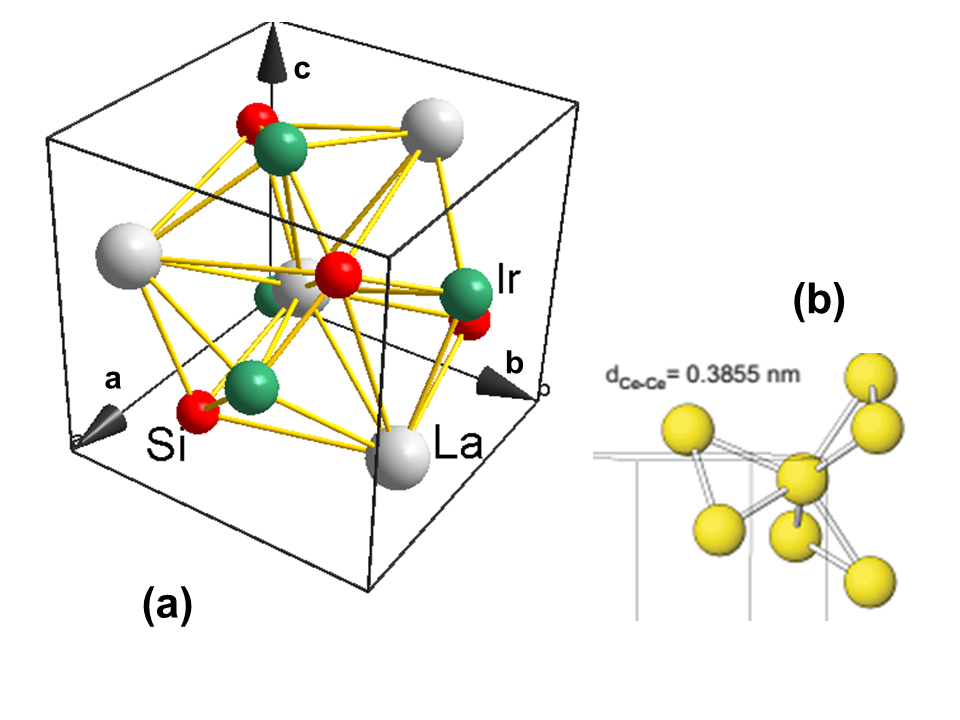}
\end{center}
\caption{(Color online) LaIrSi crystal structure type in two
coordination representations, a) with ligand atoms and b) with La
next neighbors.} \label{F1}
\end{figure}

\subsection{Magnetic, transport and thermal measurements}

The temperature dependent magnetization was measured employing a
Cryogenic superconducting quantum interference device (SQUID)
magnetometer (S700X) at temperatures from 0.3 to 2\,K with a
$^3$He-insert and from 1.8 K to room temperature with standard
$^4$He variable temperature insert and as a function of field up to 7\,T.

Electrical resistivity measurements were performed employing a standard four probe
configuration using an a.c. measurement method. Contact wires were 
made of gold, with a diameter of 50 $\mu$m respectively. 
Measurements were carried out down to 350\,mK and magnetic fields up to 9\,T.

Specific heat measurements were carried in a PPMS system 
using a He$^3$ inset to reach 400\,mK applying the relaxation 
time method below 20\,K. For this purpose, samples were prepared as a cuboid with a base of up to 2.5 mm times 2.5 mm. In general, 
a sample mass between 1\,mg to 200\,mg were mounted on the 
sample stage and attached with Apiezon N to the platform. 
Between $T=20$\,K and 
room temperature the  He$^3$ inset was removed. 

\section{Experimental results}

\subsection{Magnetic Properties}
\subsubsection{Susceptibility}

\begin{figure}[tb]
\begin{center}
\includegraphics[width=19pc]{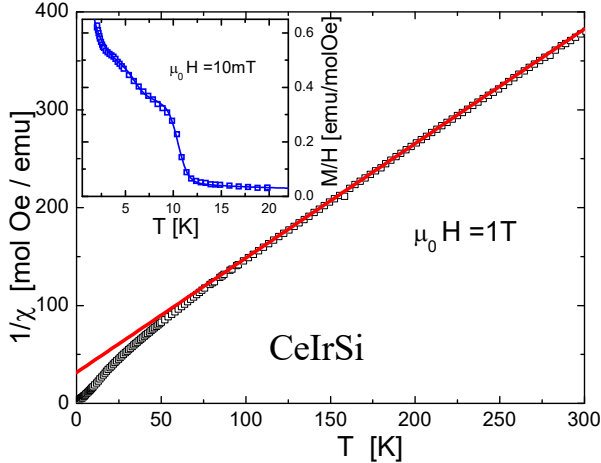}
\end{center}
\caption{(Color online) Inverse susceptibility, continuous curve represents the fit according to
Eqn.(1). Inset: analysis of the suprious ferromagnetic
contribution below 10\,K, see the text.} \label{F2}
\end{figure}

The high temperature dependent magnetic susceptibility (defined as $\chi=M/H$) in an applied field of 
1\,Tesla was measured between $T=2$\,K and
room temperature. The results are properly described in terms of the Curie-Weiss (CW) law, including a temperature independent contribution, $\chi_p$:
\begin{equation}
\chi(T)=C/(T+\theta)+\chi_P
\end{equation}

Here the Curie constant $C\propto \mu_{eff}^2$, $\mu_{eff}$ is
the effective magnetic moment, and $\theta_P$ the paramagnetic
Curie-Weiss temperature.

The Curie-Weiss law allows to analyze the magnetic susceptibility $1/\chi$ in the paramagnetic temperature range. A least squares fit for $T \geq 75$\,K according to Eq(1) is shown as a solid curve in
Fig.~\ref{F2}, revealing that $\mu_{eff} = 2.53 \mu_B$, 
in accordance to the of value of a free Ce$^{3+}$ ion. The
paramagnetic Curie temperature was derived as $\theta_P=
-21$\,K, suggesting antiferromagnetic (AFM) interactions among
the Ce$^{3+}$ ions, whereas the Pauli-like contribution was found
to be quite small, $\chi_P = 1.3 \times 10^{-4}$\,emu/mol Oe.
These values are in good agreement with Ref. \cite{Monatshefte} 
that reports a similar downwards curvature of $1/\chi$, with $\mu_{eff} = 2.56(2)
\mu_B$ and $\theta_P=-24(1)$\,K. The small difference to the present
results seems to be a result of the inclusion of the Pauli susceptibility $\chi_P$ in the data evaluation.

Below $60$\,K the inverse susceptibility significantly deviates from Curie-Weiss behavior, referring to the 
splitting of the Ce$^3+  J=5/2$ Hund's ground state due to crystalline electric field effects. Around $T\approx 10$\,K the onset of a spurious ferromagnetic
(FM) signal becomes evident (inset, Fig.~\ref{F2}). Thus, the
measured magnetic susceptibility at low temperature is described using two contributions: 
$M/H|_{meas}= M/H|_{bulk}+ M/H|_{spur}$, where $M/H|_{bulk}= 0.33/(T-0.3)$ and $M/H|_{spur}=
0.14\times tanh(10.6-T)+0.045\times atan(6-T)+0.23$.
The  $M/H|_{bulk}(T)$ term represents the CW thermal 
dependence originated by the paramagnetic doublet GS, 
whereas $M/H|_{spur}(T)$ accounts for 
the mentioned FM contribution at $T< 11$\,K 
and a weaker one at $T\approx 6$\,K. This spurious component 
can be attributed to the formation of a CeSi$_{2-x}$ (with $x \approx 0.2$) 
solid solution \cite{CeSi2x}.

\begin{figure}[tb]
\begin{center}
\includegraphics[width=18pc]{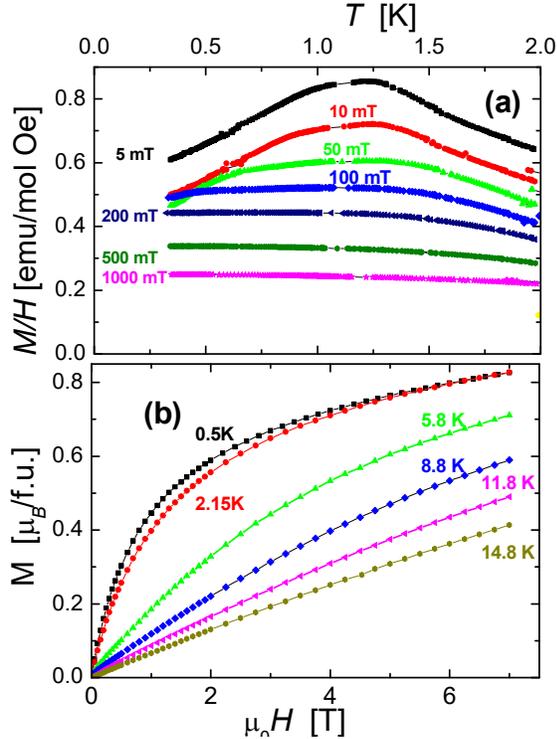}
\end{center}
\caption{(Color online) a) Low temperature magnetic 
susceptibility evaluated as $M/H(T)$ performed under 
relatively weak magnetic fields, the curves shifted 
by the effect of the FM contribution arising at $T< 
11$ \,K. b) Low temperature magnetization measurements of CeIrSi.}
\label{F3}
\end{figure}

The low temperature magnetic susceptibility ($T<2$\,K) was measured below 2\,K in applied
fields from $\mu_0 H = 5$\,mT up to 1\,T. In Fig.~\ref{F3}a details of  the temperature dependence of 
$M/H(T)$ is shown around the maximum at $T\approx 1.2$\,K. A detailed analysis of the $M/H(T)$ 
maximum reveals a weak structure; the maximum remains  
nearly constant at $T\approx 1.25$\,K until vanishing 
at $\mu_0 H \approx 200$\,mT. Additionally, a kink in M/H decreases in temperature, from $T\approx 1$\,K to  $T< 0.5$\,K at $\mu_0 H =200$\,mT. 
These features reveal a competition between two weak magnetic 
configurations which are quenched at relatively low field. 

\subsubsection{Magnetization}

The isothermal field dependence of the magnetization, measured up
to $\mu_0 H = 7$\,T is included in Fig.~\ref{F3}b. Only a slight
variation of $M$ vs $\mu_0H$ is observed between 0.5 and 2.15\,K, 
in agreement with the results presented in Fig.~\ref{F3}a. 
The paramagnetic behavior can be recognized above the $T=5.8$\,K isotherm through the collapse of 
the $M$ vs $H/T$ curves (not shown). Strictly, isothermal curves for $T\leq 8.8$\,K do not extrapolate 
to zero due to the spurious FM contribution, however its intensity is so low ( $\approx 0.01\mu_B$/f.u. at 
0.5\,K)  that it cannot be appreciated in the field scale of  Fig.~\ref{F3}b. 
According to Fig. \ref{F3}(b), the magnetization of CeIrSi at $T=0.5$~K reaches $0.82 \mu_B$ at
7\,T, with a tendency of a further increase. 
The magnetic moment of Ce$^{3+}$ in CeIrSi at low temperatures is derived from the 
respective wave function of the crystalline electric field (CEF) ground state.
For $J = 5/2$ with respect to the cubic crystal structure of CeIrSi,
a twofold ($\Gamma_7$) and a fourfold ($\Gamma_8$) degenerate state is originated.
The magnetic moment associated with the doublet is calculated 
as $M(\Gamma_7) = 0.71\mu_B$, which is derived by
 $M_{Ce^{3+}} = g_L\mu_B <\Gamma_7| J_z |\Gamma_7>
= 6/7\mu_B(1/6 |-5/2> + 5/6|3/2>) = 0.714 \mu_B$/Ce-ion.
The magnetic moment related to the quartet $\Gamma_8$, however, is much larger.
In conjunction with the so-called Van Vleck contribution, i.e., 
the non-diagonal element $<\Gamma_7| J_z |\Gamma_8>$, the 
difference between  the experimental data and the CEF magnetic moment 
might be explained.

\subsection{Electrical Resistivity}

\begin{figure}[tb]
\begin{center}
\includegraphics[width=20pc]{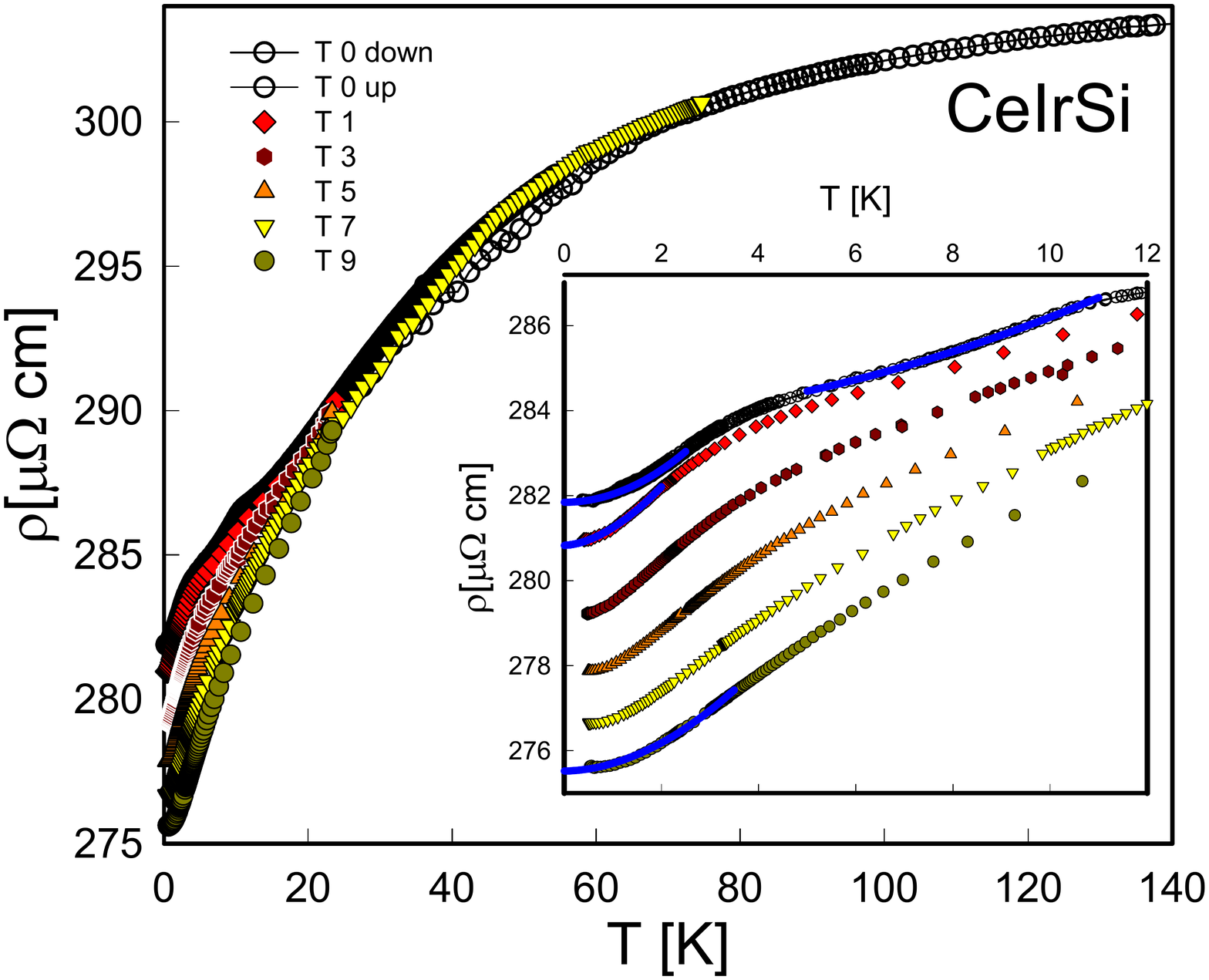}
\end{center}
\caption{(Color online) Temperature and field dependent electrical
resistivity measurements of CeIrSi. The inset shows the low  temperature
range, from 0 to 12\, K to reveal a coherent spin scattering.} \label{F4}
\end{figure}

Electrical resistivity measurements $\rho(T)$ at high 
temperature ($T>140$\,K) show a tendency to saturation 
slightly above $305\,\mu \Omega cm$ at room temperature, while from
100\,K to 15\,K the value decreases to $288\,\mu \Omega
cm$, see Fig.~\ref{F4}. Apart from the linear phonon 
contribution to $\rho(T)$, the continuous curvature can 
be associated to the progressive thermal population of 
the excited crystalline electric field level above the ground state. As 
explained above, the Ce $J = 5/2$ state in CeIrSi is split 
into a doublet and a quartet. 

A weak  kink in $\rho(T)$ around $T \approx 10.2$\,K refers to the onset
of long range magnetic order of ferromagnetic $\rm CeSi_{1.7}$.
The rather small signal change, however,  indicates 
just a low volume fraction, which, in addition, becomes fully suppressed
by rising magnetic fields (compare Fig. \ref{F4}).

With decreasing temperature, the system enters into a coherent spin
fluctuation regime below about 2.5\,K, as evidenced by  a $T^2$ temperature dependence
of $\rho(T)$ (solid lines, inset, Fig. \ref{F4}). This regime 
appears to be quite robust with respect to distinct changes observed in both $C_p(T)$ and $\chi(T)$
in this temperature range. 

By  increasing magnetic fields the Ce-spins 
get aligned along the external field direction and thus reduce the
electrical resistivity. Notably, a negative residual 
magnetoresistivity $\rho_0(H)$ at $T\to 0$  shows an almost 
linear dependence with a small ratio $\Delta \rho_0 / \Delta H \approx 0.7\,\mu\Omega cm/T$ and 
an expanding range of spin fluctuation type behaviour. 
This reveals a magnetic scattering component in $\rho_0$ 
that is reduced by increasing magnetic field. On the other hand the influence of external magnetic 
fields up to 9\,T appears to be negligible at temperatures above 60\,K

\subsection{Specific Heat}

\begin{figure}[tb]
\begin{center}
\includegraphics[width=18pc]{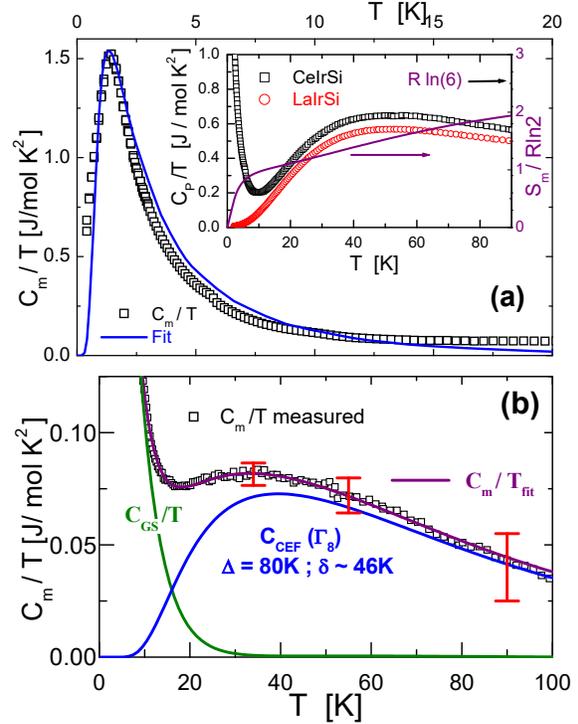}
\end{center}
\caption{(Color online) a) Low temperature magnetic contribution $C_m(T)/T$ up to 20\,K and (continuous curve) comparison with theoretic prediction for a trillium-type lattice \cite{trilium}. Inset: (left 
axis) measured specific heat $C_P(T)/T$ of CeIrSi and LaIrSi in a temperature range up to 90\,K and (right 
axis) entropy variation $S_m(T) $ normalized to a doublet GS - R$\ln(2)$. b) High temperature $C_m(T)/T
$ of CeIrSi showing the analysis for GS and excited $\Gamma_8$ quartet contributions (see the text). Error bars are representative of the uncertainty of the $C_P(T)$ measurements.} \label{F5}
\end{figure}

Specific heat measurements $C_P(T)$ provide a deeper insight on
the GS nature of CeIrSi. Zero field measurements, performed from
1.9\,K up to 80\,K are shown in the inset of Fig.~\ref{F5}a. The magnetic specific heat contribution 
$C_{m}$ was obtained by subtracting the phonon contribution $C_{ph}$
extracted from the isotypic compound LaIrSi \cite{Kneidinger}, i.e.$C_{m}=C_P - C_{ph}(\rm LaIrSi)$. At
low temperature $C_P(\rm LaIrSi)$ can be described simply by $ C_P(\rm LaIrSi)  = \gamma T + \beta T^3$ with $\gamma =
2.8$\,mJ/mol\,K$^2$ and $\beta = 0.672$\,mJ/mol\,K$^4$. In the inset of Fig.~\ref{F5}a, the specific heat is depicted for both compounds up to 90 K.

The FM transition of CeSi$_{1.8}$ at $T\approx 10$\,K, weakly present in
$\rho(T)$ measurements, is not observed in specific heat at all 
because of the small amount of the involved mass. 
The most relevant feature observed in Fig.~\ref{F5}a is the $C_m(T)/T$ maximum centered at $T^* \approx 1.5$\,K, that almost coincides in temperature with the maximum in the 
magnetic susceptibility as presented in detail in Fig.~\ref{F6}a. Notably, there is no distinct specific heat jump associated to the $C_m(T)/T$ maximum: the $T>T^*$ tail shows a long monotonous decrease. A 
comparison with theoretical predictions for a trillium-lattice system of spin ice type behavior, studied using  Monte Carlo calculations \cite{trilium}, is included as a solid curve in  Fig.~\ref{F5}a, after scaling the 
respective $C_m/T^*$ values. Deviations from the measured thermal dependence can be due to the finite number (six) unit cells, that does not reproduce the continuous spectrum of excitations
observed in the real system. At high temperature ($T\geq 10$\,K) there is an incipient contribution of the  excited CEF levels, not included into the model.

An analysis of the GS and excited CEF levels contributions to specific heat up to 100\,K is presented in Fig.~\ref{F5}b as $C_{m}(T)/T = C_{GS}/T + C_{CEF}/T$. As mentioned in subsection III-A, in a cubic 
symmetry the CEF splits the six fold degenerate $J=5/2$ state into a doublet ($\Gamma_7$) and a quartet  ($\Gamma_8$). Except for bcc structures, the former is the GS and the later 
the excited one, centered at the 
energy $k_B \Delta$. The $C_{CEF}(T)$ dependence is usually described by a standard Schottky type 
anomaly which, for such a level spectrum, reaches a maximum value of $C_{Sch}(T_{max}) = 6.3$\,J/mol K. This is 
not the case for CeIrSi because $C_{Sch}(T_{max})\approx 4.2$\,J/mol. This flattening of the anomaly can be attributed to a significant broadening of the excited $\Gamma_8$ quartet due to hybridisation of local and itinerant states. A simple approach to that scenario can be done by mimicking the mentioned broadening with a symmetric level distribution $\pm \delta_i$ around the nominal energy $\Delta$, using the formula:

\begin{equation}
C_{\rm CEF}=\rm R \Sigma_i A_{i}
[(\frac{\Delta \pm \delta_{i}}{2T})/2\cosh(\frac{\Delta\pm\delta_{i}}{2T})]^2
\end{equation} 
 
where R is the gas constant and $A_{i}$ a factor that accounts for the weight of each level. To describe some Lorentian distribution for the density of states distribution, the values: $A_{1}= 1/2$, $A_{2}= 1/4$ 
and $\delta_1=2 \times \delta_2$ are chosen. The model curve is compared with the experimental data in Fig.~\ref{F5}b, obtaining a very good fit up to 100\,K. The extracted values are $\Delta = 80$\,K and $
\delta_1 = 46$\,K. The later allows an estimation of the effective $\Gamma_8$ broadening and consequently a scale for the Kondo energy. Such a rough evaluation of the parameters describing quartet energy and broadening confirms the pure doublet character of the GS, which is one of the main requirements for the following discussion concerning the low temperature behavior of this compound. 

\begin{figure}[tb]
\begin{center}
\includegraphics[width=18pc]{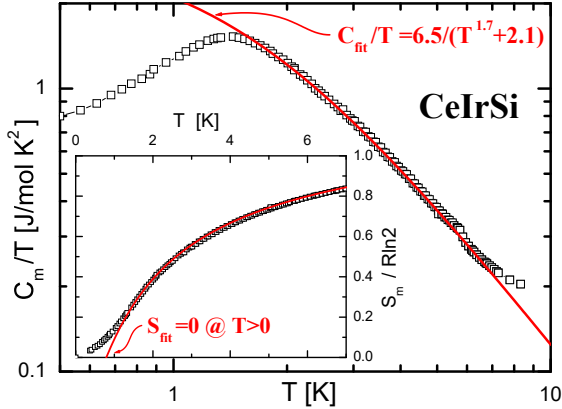}
\end{center}
\caption{(Color online) Double logarithmic representation
showing the thermal dependence of $C_m(T)/T$ above the maximum and the
corresponding power law fit (solid curve). Inset: thermal dependence of the Entropy
compared with the extrapolation of $S_{fit}(T)$ to zero (solid line) at finite temperature.}  \label{F7}
\end{figure}

In order to check the proper distribution of respective level weights, the corresponding entropy gain, extracted from this levels scheme, was computed up to room temperature, where the expected $\Delta 
S_m(\Gamma_8) = R\,ln(6/2)$ is asymptotically reached. However, by applying  this entropic analysis to the GS doublet, an excess of about 10\% of entropy was detected for the power law function describing the  $C_{GS}(T)/T$ contribution at high temperature. 
This excess can be attributed to the fact that the power law  function: $C_{GS}(T)/T = 6.5/(T^{1.7} + 2.1)$ used to describe the measured  $C_{m}(T)/T$ data dependence below about 7\,K (see for details the discussion in Subection IV-A and Fig.~\ref{F7}), does not describe the actual density of state of the physical system at higher temperature. To leave out this deviation, we have introduced a cut-off at about 20\,K, with the purpose to progressively suppress the high temperature tail. This objective was reached by subtracting a Schottky type anomaly which has a continuous increase up to a characteristic energy and a high temperature tail approaching the power low temperature dependence. The thermal energy of this cut-off was tuned such to reach the expected value of $S_{GS}= Rln2$ at high temperature.   

\section{Discussion}

\begin{figure}[tb]
\begin{center}
\includegraphics[width=18pc]{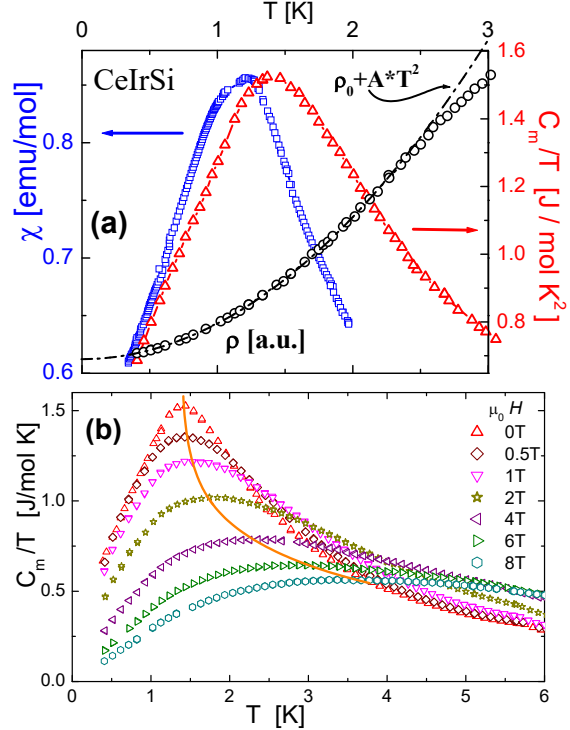}
\end{center}
\caption{(Color online) a) Comparison between $\chi(T)$,
$C_m(T)/T$ and $\rho(T)$ at $T<3$\,K. b) Specific heat dependence of CeIrSi in
magnetic field up to $\mu_0H=8$\,T. Continuous curve: guide to the eyes tracing the $C_m(T,H)/T$ maxima.} \label{F6}
\end{figure}

A comparison between $\chi(T)$, $C_m(T)/T$ and $\rho(T)$ is
presented in Fig.~\ref{F6}a. Notably both $\chi(T)$ and $C_m(T)/T$ maxima, centered between $T_{max} =1.3$ and 1.5\,K respectively, occur within the range
at which $\rho(T)\propto T^2$. Together
with the lack of a $C_m(T)/T$ jump, this behavior excludes $\chi(T)$
and $C_m(T)/T$ maxima as due to a standard phase transition.
To gain insight into the magnetic character of the GS, we have
performed specific heat measurements under magnetic fields up to 
$\mu_0 H=8$\,T, see Fig.~\ref{F6}b. The maximum of
$C_m(T,H)/T$ decreases and slightly shifts to higher temperature up to $\mu_0 H\approx 2$\,T. 
The solid line in Fig.~\ref{F6}b describes the evolution of the maximum, which broadens 
once the applied field starts to polarize the GS spins above $\mu_0H \approx 4$\,T.

\subsection{Entropy trajectory and magnetic frustration}

In order to analyze the nature of the low temperature anomaly presented in Fig.~\ref{F6}, one may compare this behavior with similar ones observed in
other intermetallics  \cite{atoms}. A common feature of those systems is the power law thermal dependence of $C_m(T)/T$ above its maximum. In Fig.~\ref{F7} this feature is verified in a double 
logarithmic representation, where the measured values are accounted for by a modified power law,
$C_{fit}/T = 6.5/(T^{1.7}+2.1)$. Such a thermal dependence is comparable with that observed in compounds recognized as frustrated systems \cite{atoms}. Differently from those recognized as spin 
glasses, with a $C_m(T) \propto 1/T^2$ tail at high temperature, one notices that the present fitted power law dependence holds up very close to the maximum with a clearly different exponent: 1.7, instead of 3 for 
the spin glass in a $C_m(T)/T$ representation. Furthermore, the $\rho \propto T^2$ dependence observed in this compound is not the expected for a spin glass \cite{Mydosh}

In frustrated systems, the $C_m(T)/T$ maximum was associated to the
temperature at which the thermal trajectory, represented by $C_{fit}(T>T_{max})/T$ in Fig.~\ref{F7} as a solid curve, changes because of thermodynamic constraints. If $C_m(T)/T$ followed the trajectory 
described by $C_{fit}(T<T_{max})/T$ (solid red line in Fig.~\ref{F7}) it would reach unphysical values at $T\to 0$. As a consequence, the entropy evaluated as $S_{fit} = \int C_{fit}/T dT$ would exceed the 
available degrees of freedom (R$\ln2$) for a doublet ground state. 

An alternative description can be done analyzing the actual trajectory of the entropy shown in the inset of 
that figure. There one can see that, if the $T>T_{max}$ values of  $S_{fit}$ are scaled with measured $S_m(T>T_{max})$, then  $S_{fit} \to 0$ at  $T>0$ that is not allowed by thermodynamics. 
Notice that in the inset of  Fig.~\ref{F7}, the high temperature value of $S_{fit}$ is referred to R$\ln2$ because of the scaling procedure. In such scenario, the Nernst postulate imposes $S_m(T)\to 0$ at $T = 
0$, undergoing an inflection point where this sort of 'entropy bottleneck' occurs \cite{atoms}.  
This fact indicates that the $C_m(T)/T$ maximum is driven by a thermodynamic constraint instead of  classical magnetic interactions effect. 

Divergent power laws for the $C_m(T)/T$ dependence are a characteristic of these frustrated systems,  
because low energy magnetic excitations strongly accumulate at $T\to 0$. This is due to the fact that no 
order parameter, able to reduce the GS degeneracy, can develop. 
Since entropy accumulation cannot exceed the 
available degrees of freedom provided by the doublet GS, the system is forced into an alternative 
minimum of the free energy \cite{JLTP18}. Since this 
transition occurs in a continuous way, no discontinuity (or jump) 
is observed in  $C_m(T)/T$, whilst such a discontinuity is observed in $\partial C_m/\partial T$, i.e. the 
third derivative of the free energy. Even the faint structure observed in the magnetic susceptibility between 1 and 1.2\,K may reveal a competition between two broad minima in the free energy which are 
blurred out by moderate magnetic field.  The origin of such entropy bottleneck can be understood in the context of magnetic frustration of magnetic interactions due to a peculiar geometrical configuration, like 
the 3D network of corner-sharing triangles presented in Fig.~\ref{F1}, which mimics a trillium flower. 

A relatively large paramagnetic Curie-Weiss temperature compared with the corresponding ordering temperature is frequently used to define a frustration parameter: $f = \frac{\theta_p}{T_{ord}}$ 
\cite{Ramirez}. 
This heuristic criterion reflects the decrease of $T_{ord}$ in respect to the expected values evaluated within a mean field approximation. In the case of CeIrSi, the $\chi(T)$ and $C_m(T)/T$ maxima around 
1.3~K are more than one order of magnitude smaller than $\theta_p = -21$~K, revealing a ratio $f > 10$ that  hints to a frustration scenario for the magnetic moments.

\section{Conclusions}

The peculiar trillium type crystalline structure of CeIrSi provides the possibility to study the effects of magnetic  frustration in a 3D Ce-lattice.  The entropy driven character of the anomaly, observed around 
1.3\,K, is deduced from the divergent power law dependence of $C_m(T)/T$. At that temperature, the   entropy is constrained to change trajectory in order to not overcome the $S_m=Rln2$ limit imposed by the 
Nernst postulate. Notably, there is a number of compounds showing similar $C_m(T)/T$ anomalies, 
followed by very similar power law dependencies at higher temperature \cite{atoms}, all of them related to underlying frustration features. The present compound, with trillium type structure, exhibits the same spin-
ice character than pyrochlore structured ones, c. f. Dy$_2$Ti$_2$O$_7$, suggesting that magnetically frustrated paramagnets slide into an alternative free energy minimum driven by entropy constraints.    

The $\rho \propto T^2$ dependence, along the range where $\chi(T)$ and $C_m(T)/T$ maxima show up, reveals that the nature of random interactions occurring in a magnetically frustrated configuration clearly differs from a spin glass scenario. Despite some common features related to disordered interactions, like the effect of magnetic field observed in the $M(T)/H$ and $C_m(T)/T$ dependence around respective maxima may arise, the electron-spin scattering coherence of frustrated systems reveals distinct differences 
between dynamic and frozen landscapes. 


\end{document}